\documentclass[12pt,notitlepage,a4paper]{article}

\pdfoutput=1

\usepackage{a4wide}
\usepackage{epsfig}
\usepackage{color,graphicx}
\usepackage{cite}
\usepackage{amssymb,amsmath}
\newcommand{\be}{\begin{equation}}
\newcommand{\ee}{\end{equation}}
\newcommand{\bea}{\begin{eqnarray}}
\newcommand{\eea}{\end{eqnarray}}

\begin{document}

\begin{center}  

\vskip 2cm 

\centerline{\Large {\bf An $\mathcal{N}=1$ Lagrangian for the rank $1$ $E_6$ superconformal theory}}

\vskip 1cm

\renewcommand{\thefootnote}{\fnsymbol{footnote}}

   \centerline{
    {\large \bf Gabi Zafrir${}^{a, b}$} \footnote{gabi.zafrir@unimib.it}}

\vspace{1cm}
\centerline{{\it ${}^a$ Kavli IPMU (WPI), UTIAS, the University of Tokyo, Kashiwa, Chiba 277-8583, Japan}}
\centerline{{\it ${}^b$ Dipartimento di Fisica, Universit\`a di Milano-Bicocca \& INFN, Sezione di Milano-Bicocca,}}
\centerline{{\it I-20126 Milano, Italy}}
\vspace{1cm}

\end{center}

\vskip 0.3 cm

\setcounter{footnote}{0}
\renewcommand{\thefootnote}{\arabic{footnote}}   
   
      \begin{abstract}
     
		We propose that a certain $4d$ $\mathcal{N}=1$ $SU(4)$ gauge theory flows in the IR to the rank $1$ $\mathcal{N}=2$ strongly coupled SCFT with $E_6$ global symmetry and $25$ free chiral fields. This proposal is tested by comparing various RG invariant quantities, notably, anomalies and the superconformal index. We discuss the generalization to $\mathcal{N}=1$ $SU(2n+2)$ gauge theory models flowing in the IR to the $R_{(2,2n+1)}$ family of strongly coupled SCFTs plus free fields.
		
      \end{abstract}

\tableofcontents

\section{Introduction}

In the space of quantum field theories a special place is reserved to conformal theories. This comes from our perception of a quantum field theory as an RG flow between a UV theory and an IR theory that are thought to have conformal symmetry. As a result conformal theories provide the starting and end points of RG flows, and their space has potential to teach us much about the space of quantum field theories in general. Thus, the study of conformal theories, especially interacting ones, has received much attention.

In four dimensions it is possible to build interacting Lagrangian conformal theories if enough supersymmetry is present. The simplest example is the maximally supersymmetric Yang-Mills theory which is conformal regardless of coupling, and so in particular is also conformal at weak coupling. This can be generalized also to gauge theories with only $\mathcal{N}=2$ or even just $\mathcal{N}=1$ supersymmetry though then various conditions on the matter content must be satisfied\cite{BT,LS,GKSTW}. However, it has been realized for quite some time that there are also strongly interacting SCFTs, that do not have a weak coupling limit and so no Lagrangian description. The prime examples are the Minahan-Nemeschansky $E$-type theories \cite{MN}, though many additional examples have been found, see for instance \cite{AW,Gai,CD,CD1}.

In recent years, however, it has become apparent that these seemingly non-Lagrangian theories may still posses a Lagrangian albeit one preserving less supersymmetry or symmetry. These constructions usually fall into one of two types. One is to start with an $\mathcal{N}=1$ Lagrangian theory which flows to the strongly interacting $\mathcal{N}=2$ SCFT at the IR, potentially with the addition of decoupled fields. This has been used to realize Lagrangians of this type for some $\mathcal{N}=2$ SCFTs, mostly of the so called Argyres-Douglas type \cite{MS,MS1,AMSad,ASS,BGad,MNS}. A closely related construction is to consider $\mathcal{N}=1$ Lagrangians that involve gauging of symmetries that appear only at strong coupling, and so are not actually visible in the Lagrangian\cite{GRW,RVZ,AMS,KRVZ}. As a result, these are not completely Lagrangian in the usual sense, but can still be used to perform various computations. 

An alternative construction is to seek an $\mathcal{N}=1$ Lagrangian theory that sits on the same conformal manifold as the $\mathcal{N}=2$ SCFT\cite{RZCM}. While these strongly coupled $\mathcal{N}=2$ SCFTs have no $\mathcal{N}=2$ preserving conformal manifold, some of them have an $\mathcal{N}=1$ only preserving conformal manifold. It is then possible that there exist $\mathcal{N}=1$ Lagrangian theories that sit on the same conformal manifold as the non-Lagrangian SCFT. The strongly coupled SCFT can then be thought of as a strong coupling limit of the Lagrangian theory, in the same vein as strong coupling points in the maximally supersymmetric Yang-Mills theory, albeit here with enhanced supersymmetry that is entirely accidental from the viewpoint of the Lagrangian theory.

An interesting question then is are there really theories with no Lagrangian description also of these types? Currently we know of such Lagrangians just for a small subset of the known strongly coupled $\mathcal{N}=2$ SCFTs. As a first step it seems interesting to try to see if this can be extended to more cases. In this article we shall consider this problem, attempt to develop methods to conjecture such Lagrangians, and illustrate them by constructing Lagrangians for a family of strongly interacting SCFTs. Specifically, we shall consider the rank $1$ Minahan-Nemeschansky $E_6$ theory, one of the simplest and most well-known of the non-Lagrangian theories. Using various considerations, we shall conjecture a simple $\mathcal{N}=1$ Lagrangian theory which we claim may flow in the IR to this strongly coupled $\mathcal{N}=2$ SCFT and several decoupled chiral fields. We then test this conjecture with various consistency checks, notably through the comparison of various coupling invariant quantities. We should note that there is already a known Lagrangian for this theory, although one involving the gauging of a symmetry not manifest in the Lagrangian\cite{GRW}. The Lagrangian we suggest here does not involve this complication.     

Following that we shall generalize the Lagrangian so as to obtain potential Lagrangians also for the $R_{2,2n+1}$ family of SCFTs \cite{CD}. As we will show the Lagrangians pass various non-trivial consistency checks. These Lagrangians allow us to also construct Lagrangians for many other $\mathcal{N}=2$ SCFTs, built by conformally gauging part of the global symmetry of the $R_{2,2n+1}$ SCFTs. Notably, as the rank $1$ $E_6$ SCFT is the same as $T_3$, this includes SCFTs generated via the compactification of the $A_2$ $(2,0)$ theory on generic Riemann surfaces\cite{Gai}. However, as the global symmetry of the $\mathcal{N}=1$ Lagrangian theory only enhances to that of the $\mathcal{N}=2$ SCFT at the IR, some of these Lagrangians may involve gauging of symmetries not manifest in the Lagrangian.

The structure of this article is as follows. We start in section \ref{sec:pre} with some preliminary discussion, outlining the general strategy adopted in this paper. We then introduce and study the proposed Lagrangian for the rank $1$ MN $E_6$ theory in section \ref{sec:E6}. This is generalized to the $R_{2,2n+1}$ families of SCFTs in section \ref{sec:Rodd}. The appendix summarized various properties of the $R_{2,2n+1}$ family of theories that would prove useful in this article.

\section{Preliminaries}
\label{sec:pre}

We are interested in the problem of finding $\mathcal{N}=1$ Lagrangian theories that upon a deformation by relevant or marginal operators become strongly coupled $\mathcal{N}=2$ SCFTs. The strongly coupled $\mathcal{N}=2$ SCFTs fall into several classes. First, there are the ones of Argyres-Douglas type \cite{ArD,APSW}. These contain operators with fractional dimensions, and usually also operators whose dimension is smaller than $1$. Then there are also non-Lagrangian theories like the MN $E_6$, $E_7$ and $E_8$ theories whose spectrum seems to contain only operators of integer dimensions. These latter type of theories are usually related to one another and Lagrangian superconformal gauge theories via a conformal gauging of part of their global symmetry, a phenomena originally discovered in \cite{ArS}. As such these are quite similar to Lagrangian $\mathcal{N}=2$ SCFTs. We will be mostly interested in finding Lagrangians for theories belonging to this class.

Let us consider a $4d$ $\mathcal{N}=1$ weakly coupled Lagrangian theory. For concreteness, we shall assume it is described by vector multiplets in the adjoint of a gauge group $G$, not necessarily simple, and chiral fields in some representation of $G$. The gauge interactions may be either asymptotically free, in which case they are weakly coupled at high energies, or confomal, in which case the couplings can be tuned so that the theory is weakly coupled. They can also be IR free, but these are expected to have trivial dynamics in the IR and so will not be considered here\footnote{One can consider also more intricate scenarios where some groups are asymptotically free and some are IR free at high energies, but the beta function of the IR free groups eventually vanishes due to the strong coupling dynamics of the asymptotically free groups, see for instance \cite{KRVZ1}. We shall not consider these cases here.}. At the weak coupling point, the dimension of all operators is integer, and the dimension of the chiral fields is $1$. The spectrum then looks quite similar to the strongly coupled $\mathcal{N}=2$ SCFTs we are interested in. We can next consider what happens as we go to lower energies. If the theory is asymptotically free than we expect the dimension of at least some of the operators to go down\footnote{This follows from the NSVZ beta function. If it is initially negative, as it is for asymptotically free theories, then in order for it to vanish, at least one field must develop a negative anomalous dimension.}. As a result we generally expect to end with theories that have operators with fractional dimensions, and indeed most known Lagrangian $\mathcal{N}=1$ asymptotically free gauge theories flowing to interacting strongly coupled $\mathcal{N}=2$ SCFTs, flow to SCFTs of the Argyres-Douglas type.

If we wish to find Lagrangians for the theories of the second type we discussed, we must somehow keep all the dimensions integer. The simplest way to do that is to have no flow, that is to have a conformal theory. This gives the type of Lagrangians considered in \cite{RZCM}. However, not all SCFTs of this type have an $\mathcal{N}=1$ preserving conformal manifold so this method cannot give Lagrangians for all cases.

Instead we must consider asymptotically free cases such that the dimensions are forced to be integers at the end of the flow. The simplest scenario is to start with a theory where the minimal BPS operator has dimension $2$, but we end up with dimension $1$ BPS operators. That is, the theory we end up with is the strongly coupled SCFT plus free fields. Here we shall mostly concentrate on BPS operators as we can determine the IR dimension of these operators, under some assumptions, even-though the examples we are about to consider involve flows to strong coupling. Specifically, we can determine the dimension of these operators from their R-charge, under the assumption that the theory flows to a superconformal fixed point with that R-symmetry as the superconformal one. In $4d$ the relation is:

\be
\Delta = \frac{3 R}{2},
\ee
for a scalar operator of superconformal $U(1)_R$ charge $R$ and dimension $\Delta$.  

To achieve such a scenario we shall adopt the following strategy. We shall construct an $\mathcal{N}=1$ Lagrangian gauge theory with a $U(1)_R$ symmetry under which the chiral fields have R-charges of either $\frac{2}{3}$ or $\frac{1}{3}$. We then seek to force this $U(1)_R$ symmetry to be the superconformal one by using deformations. In addition, if the theory is chosen so that there are no gauge invariants possible using an odd number of chiral fields with R-charge $\frac{1}{3}$, then this forces all BPS operators to have integer dimensions. Quadratic gauge invariants made from the operators with R-charge $\frac{1}{3}$ become free fields that should decouple.

Next, we need to consider how to search for such models. Here we shall borrow the strategy used in \cite{RZCM}. Specifically, we use the conformal anomalies of the strongly coupled SCFT to help narrow the search. The conformal anomalies are determined by the anomalies in the $U(1)_R$ symmetry, which in turn we can calculate immediately as by assumption the matter spectrum contains only gauge fields and chiral fields with R-charge $\frac{2}{3}$ and $\frac{1}{3}$. Denoting the number of vectors as $n_v$, the number of chirals with R-charge $\frac{2}{3}$ as $n^{(\frac{2}{3})}_c$ and the number of chirals with R-charge $\frac{1}{3}$ as $n^{(\frac{1}{3})}_c$, we have:

\be
a = \frac{1}{48}(9 n_v + n^{(\frac{2}{3})}_c - n^{(\frac{1}{3})}_c) \; , \; c = \frac{1}{48}(6 n_v + 2 n^{(\frac{2}{3})}_c + n^{(\frac{1}{3})}_c) .
\ee

This number must match the central charges of the SCFT plus the free fields. The strategy now is as follows. Choose a strongly coupled SCFT. Look for a gauge theory such that the numbers $n_v, n^{(\frac{2}{3})}_c$ and $n^{(\frac{1}{3})}_c$ are consistent with the $a$ and $c$ central charges of the strongly coupled SCFT plus the free fields, which are fixed to be all the quadratic gauge invariants built purely from $n^{(\frac{1}{3})}_c$. The gauge theory is further constrained by additional demands, like that the needed R-symmetry be anomaly free. This leads to a rather constrained game, involving a finite number of possibilities. All that remains is the lengthy work of going through all of them, but if a Lagrangian of this type exists for the chosen SCFT, then we are guaranteed to find it eventually.   

As an example let's consider applying this procedure to the rank $1$ $E_6$ theory, which is one of the simplest SCFTs of this type. The $a$ and $c$ central charges of this theory are \cite{TA}:

\be
a_{E_6} = \frac{41}{24} \quad , \quad c_{E_6} = \frac{13}{6}  \label{acE6}.
\ee

From these we find that:

\be
n^{(\frac{1}{3})}_c = 4 n_v - 20 \; , \; n^{(\frac{2}{3})}_c = 62 - 5 n_v .
\ee

The value of $n_v$ is constrained to be the sum of dimensions of simple groups, which for small values of $n_v$ is quite limiting. Going over many cases, we find a solution obeying all of the mentioned constraints. Specifically, the solution is:

\be
n_v = 15 \; , \; n^{(\frac{1}{3})}_c = 40 \; , \; n^{(\frac{2}{3})}_c = -13 .
\ee 

The negative number for $n^{(\frac{2}{3})}_c$ might be alarming at first, but as we shall see it just follows from the fact that there are more free fields at the IR than fields with R-charge $\frac{2}{3}$ in the UV. The specific theory we consider is an $SU(4)$ gauge theory with five chiral fields in the fundamental, five chiral fields in the antifundamental and two chiral fields in the antisymmetric. The fundamental and antifundamental chiral fields have R-charge of $\frac{1}{3}$, and the antisymmetric chiral fields have R-charge of $\frac{2}{3}$. These R-charges are forced by a superpotential, on which we shall elaborate in greater detail in the next section. The fundamental and antifundamental chiral fields indeed give 40 chiral fields with R-charge $\frac{1}{3}$. From them we can make 25 quadratic gauge invariants, which are just the mesons. Thus, at the end of the flow we expect to have a potentially interacting piece and 25 free chiral fields. The effective number of chiral fields with R-charge $\frac{2}{3}$ in the part without the 25 free fields is $12-25=-13$, as needed. This ensures that the anomalies of the part after the removal of the free fields matches the anomalies of the rank $1$ $E_6$ SCFT. Another important requirement is that the gauge theory does not suffer from gauge anomalies and the R-symmetry we need is anomaly free. The former follows as the theory is charge conjugation invariant and the latter follows as:

\be
4+5\times \frac{1}{2} \times (-\frac{2}{3}) +5\times \frac{1}{2} \times (-\frac{2}{3}) +2\times 1 \times (-\frac{1}{3}) = 0.
\ee 

Finally, we might worry that due to the presence of gauge variant operators with R-charge $\frac{1}{3}$, there might be gauge invariant operators with fractional dimensions. However, it is straightforward to show that all gauge invariants have R-charges which are multiples of $\frac{2}{3}$. This follows as $SU(4)$ has a $\mathbb{Z}_4$ center symmetry, the $\mathbb{Z}_2$ subgroup of which acting non-trivially only on the fundamental and antifundamental matter. Thus, all $SU(4)$ gauge invariants must contain an even number of these fields, and as these are the only fields with R-charge $\frac{1}{3}$, all gauge invariants must have R-charges which are multiples of $\frac{2}{3}$. 

Next, we shall perform a more in-depth study of this theory. We shall consider its dynamics, and whether the R-symmetry we use can be made to be the superconformal R-symmetry. We shall also perform various consistency checks by matching flow invariant quantities notably, anomalies and the superconformal index. These will also allow us to give evidence that the global symmetry enhances to $E_6$. 

\section{Dual for the rank $1$ $E_6$ theory}
\label{sec:E6}

Let us analyze in detail the proposed dual of the rank $1$ $E_6$ theory that we previously suggested. The matter content consists of an $SU(4)$ gauge theory with two chiral fields in the antisymmetric representation, five chiral fields in the fundamental and five chiral fields in the antifundamental. This theory has non-anomalous global symmetry given by $SU(5)^2 \times SU(2)\times U(1)_b \times U(1)_c \times U(1)_R$, where the charges of the various fields are summarized in the table below. Particularly, it has a non-anomalous R-symmetry under which the antisymmetric chirals have free R-charge $\frac{2}{3}$ and the fundamental ones have R-charge $\frac{1}{3}$.

\begin{center}
	\begin{tabular}{|c||c|c|c|c|c|c|c|}
		\hline
		Field & $SU(4)$ & $SU(5)_F$ & $SU(5)_{\overline{F}}$ & $SU(2)$ & $U(1)_b$ & $U(1)_c$ & $U(1)_{R}$\\
		\hline
		$A$ & $\boldsymbol{6}$ & $\boldsymbol{1}$ & $\boldsymbol{1}$ & $\boldsymbol{2}$ & $0$ & $5$ & $\frac{2}{3}$ \\
		
		$F$ & $\boldsymbol{4}$ & $\boldsymbol{5}$ & $\boldsymbol{1}$ & $\boldsymbol{1}$ & $1$ & $-2$ & $\frac{1}{3}$ \\
		
		$\overline{F}$ & $\boldsymbol{\overline{4}}$ & $\boldsymbol{1}$ & $\boldsymbol{5}$ & $\boldsymbol{1}$ & $-1$ & $-2$ & $\frac{1}{3}$ \\
		\hline
	\end{tabular}
\end{center}

Without a superpotential this R-charge is not the superconformal one. This occurs due to mixing with $U(1)_c$. There is no mixing with $U(1)_b$ as the matter fields come in equivalent representation with opposite charges. We can determine the mixing using the technique of a maximization \cite{Amax}. Specifically, we define the trial R-symmetry $U(1)^{trial}_R = U(1)_R + \gamma U(1)_c$ and compute:

\bea
Tr \left((U(1)^{trial}_R)^3 \right) & = & 15 + 12 (-\frac{1}{3} + 5\gamma)^3 + 40 (-\frac{2}{3} - 2\gamma)^3, \label{Rtraces} \\ \nonumber Tr \left(U(1)^{trial}_R \right) & = & 15 + 12 (-\frac{1}{3} + 5\gamma) + 40 (-\frac{2}{3} - 2\gamma) . 
\eea 

We then use

\be
a = \frac{3}{32} \left(3 Tr(U(1)^3_R) - Tr(U(1)_R)\right) \quad , \quad c = \frac{1}{32} \left(9 Tr(U(1)^3_R) - 5 Tr(U(1)_R)\right) \label{aandc}
\ee

to compute the central charges, notably $a$. The statement of a maximization is that the superconformal R-symmetry is the one that maximizes $a$. We can use this to fix the coefficient $\gamma$ that determines the mixing between $U(1)_R$ and $U(1)_c$. We find that $\gamma = -\frac{31 - \sqrt{1669}}{177} \approx -0.056$. 

This result should give the superconformal R-symmetry in the IR. However, this may fail if there are accidental $U(1)$ symmetries arising in the IR, that can then mix with the R-symmetry \cite{KPS}. It is generally very difficult to completely rule this out, but one consistency check one can do is to verify that all BPS operator dimensions, expected from their superconformal R-charge, are above the unitarity bound. Operators below the unitarity bound are inconsistent in SCFTs suggesting that the IR theory cannot be an SCFT with that superconformal R-symmetry. It is generally thought that in these cases the violating operators decouple, and become free fields along the flow. This leads to additional symmetries that then mix with the R-symmetry. Returning to the case at hand, with the R-symmetry we found all gauge invariant operators are above the unitarity bound and so it seems plausible that this theory flows to an interacting SCFT. Specifically, the mesons $F \overline{F}$ have dimension bigger than $1$. 

We next want to consider the operator $Tr (A^2 F \overline{F})$, where we form an $SU(4)$ invariant in the following manner. We can contract the two $SU(4)$ antisymmetrics to form an $SU(4)$ adjoint representation, which is an antisymmetric contraction. Similarly we can contract the fundamental and antifundamental to form an $SU(4)$ adjoint representation. Finally we can contract the two adjoints to form a singlet. The resulting operator is in the $(\boldsymbol{5}, \boldsymbol{5}, \boldsymbol{1})$ of the $SU(5)_F \times SU(5)_{\overline{F}}\times SU(2)$ global symmetry, where it is an $SU(2)$ singlet as the contraction of the two $SU(4)$ antisymmetrics is done antisymmetrically in the gauge indices and so must also be done antisymmetrically in the flavor indices so that the full contraction be symmetric. It is also a singlet under $U(1)_b$, has charge $6$ under $U(1)_c$ and R-charge $2$ under $U(1)_R$. The last two imply that this operator is relevant with respect to the superconformal R-symmetry. This comes about as $\gamma$ is negative and so this operator has superconformal R-charge smaller than $2$, meaning dimension smaller than $3$. 

 Since this operator is relevant at the fixed point, introducing it to the superpotential will initiate a flow leading to a new fixed point. The introduction of this operator breaks $U(1)_c$ and also breaks $SU(5)_F \times SU(5)_{\overline{F}}$ to the diagonal $SU(5)$. As $U(1)_c$ is broken the R-symmetry is forced to be $U(1)_R$, as there is no other $U(1)$ it can mix with, under the assumption that there are no accidental $U(1)$ symmetries arising in the IR that can mix with the R-symmetry. Under this R-symmetry there are gauge invariant fields that have the free R-charge $\frac{2}{3}$, specifically, the mesons $F \overline{F}$. These should then decouple at the IR and we expect to get $25$ free chiral fields plus a remaining fixed point\footnote{As we expect to get decoupled free fields there should also be accidental symmetries acting only on them. However, as the R-charge of these fields is locked to $\frac{2}{3}$, these will not mix with the R-symmetry.}. We claim that this remaining fixed point is the rank $1$ $E_6$ theory. We next present evidence for this claim.

\subsection*{Index}

We start by considering the superconformal index of this theory \cite{Index}. Since the index is invariant under flows, the index of the UV theory should be equal to that of the IR fixed point, assuming the symmetries were identified correctly. As the UV theory is a Lagrangian theory, it is straightforward to evaluate the index in the UV, which then must match the index of the expected IR theory, if our conjecture is to be correct. 

We are then lead to calculate the superconformal index of the UV theory. However, as we expect some fields to decouple and become free fields in the IR, it is convenient to modify the Lagrangian slightly so as to remove them. This can be done through the procedure of flipping \cite{BG}. The idea is to introduce additional singlet fields $M$ and $N$ and couple them to the mesons, which are the fields that we expect to decouple in the IR, via the superpotential $W = M F \overline{F} + N F \overline{F}$. Here $M$ and $N$ differ by their representation under the preserved $SU(5)$. The charges of the two fields under the preserved global symmetry are shown in the table below. The F term relations of these fields eliminate the mesons, $F \overline{F}$, from the chiral ring, and thus these are no longer present in the IR. This can also be seen physically as the superpotential becomes a mass term for the singlets and the mesons, once the latter decouple. This makes the index computation easier, and is also conceptually convenient as it circumvents having free fields in the IR and the additional global symmetries associated to them.   

\begin{center}
	\begin{tabular}{|c||c|c|c|c|c|}
		\hline
		Field & $SU(4)$ & $SU(5)$ & $SU(2)$ & $U(1)_b$ & $U(1)_{R}$\\
		\hline
		$M$ & $\boldsymbol{1}$ & $\boldsymbol{24}$ & $\boldsymbol{1}$ & $0$ & $\frac{4}{3}$ \\
		
		$N$ & $\boldsymbol{1}$ & $\boldsymbol{1}$ & $\boldsymbol{1}$ & $0$ & $\frac{4}{3}$ \\
		\hline
	\end{tabular}
\end{center}

We can then evaluate the index with the flip fields finding:

\bea
I & = & 1 + (p q)^{\frac{2}{3}} (\boldsymbol{3}_{SU(2)} + b^2 \boldsymbol{2}_{SU(2)} \boldsymbol{10}_{SU(5)} + \frac{1}{b^2} \boldsymbol{2}_{SU(2)} \boldsymbol{\overline{10}}_{SU(5)} + \boldsymbol{24}_{SU(5)} + 1 + b^4 \boldsymbol{\overline{5}}_{SU(5)} + \frac{1}{b^4} \boldsymbol{5}_{SU(5)}) \nonumber \\  & - & p q (\boldsymbol{3}_{SU(2)} + b^2 \boldsymbol{2}_{SU(2)} \boldsymbol{10}_{SU(5)} + \frac{1}{b^2} \boldsymbol{2}_{SU(2)} \boldsymbol{\overline{10}}_{SU(5)} + \boldsymbol{24}_{SU(5)} + 1 + b^4 \boldsymbol{\overline{5}}_{SU(5)} + \frac{1}{b^4} \boldsymbol{5}_{SU(5)}) + ... , \nonumber \\  &&
\eea
where we use the standard notations \cite{DO,RR} for the index, and $b$ is the fugacity for $U(1)_b$. The collection of terms at orders $(p q)^{\frac{2}{3}}$ and $p q$ build the adjoint of $E_6$, where the global symmetry is embedded as $U(1)_b \times SU(5)\times SU(2) \subset SU(6)\times SU(2)\subset E_6$. To see this we note that under this series of embeddings we have that $\boldsymbol{78}_{E_6} \rightarrow \boldsymbol{3}_{SU(2)} + \boldsymbol{35}_{SU(6)} + \boldsymbol{2}_{SU(2)} \boldsymbol{20}_{SU(6)} \rightarrow \boldsymbol{3}_{SU(2)} + \boldsymbol{24}_{SU(5)} + b^4 \boldsymbol{\overline{5}}_{SU(5)} + \frac{1}{b^4} \boldsymbol{5}_{SU(5)} + 1 + \boldsymbol{2}_{SU(2)} (b^2\boldsymbol{10}_{SU(5)} + \frac{1}{b^2}\boldsymbol{10}_{SU(5)})$, where in the last transition we used the decomposition $\boldsymbol{6}_{SU(6)} \rightarrow b^{\frac{2}{3}} \boldsymbol{\overline{5}}_{SU(5)} + \frac{1}{b^{\frac{10}{3}}}$. In terms of $E_6$ characters, the index reads:

\be
I = 1 + (p q)^{\frac{2}{3}} \boldsymbol{78}_{E_6} - p q \boldsymbol{78}_{E_6} + (p q)^{\frac{2}{3}}(p+q) \boldsymbol{78}_{E_6} + (p q)^{\frac{4}{3}} (2+\boldsymbol{2430}_{E_6}) + ... . \label{IndexSU4}
\ee

In theories without free field, the terms contributing to the index as $(p q)^{m}$, for $m\leq 1$, have a direct physical interpretation \cite{GB}. Specifically, for $m<1$ these are relevant operators, and for $m=1$ these are either marginal operators, which contribute positively, or conserved currents, which contribute negatively. Here the first terms then are relevant operators, and can be identified with the moment map operators expected in an $\mathcal{N}=2$ SCFT. Indeed, these are in the adjoint of the global symmetry of the expected $\mathcal{N}=2$ SCFT. The second terms then give the marginal operators minus the conserved currents. Here we see the adjoint of $E_6$ contributing negatively, which can be readily identified with the conserved currents of an $E_6$ global symmetry. This is a strong indication that the low-energy theory has an $E_6$ global symmetry, under the assumption of a superconformal fixed point. Finally we can compare with the known index of the rank $1$ $E_6$ SCFT, evaluated in \cite{E6Index}: 

\be
I = 1 + \frac{1}{\nu}(p q)^{\frac{2}{3}} \boldsymbol{78}_{E_6} + p q (\nu^3 - \boldsymbol{78}_{E_6} - 1) + \frac{1}{\nu}(p q)^{\frac{2}{3}}(p+q)(1 + \boldsymbol{78}_{E_6} - \nu^3 ) + (p q)^{\frac{4}{3}} (2\nu + \frac{1}{\nu^2}\boldsymbol{2430}_{E_6}) + ... . \label{IndexE6}
\ee

Here $\nu$ is the fugacity for the additional $U(1)$ commutant of the $\mathcal{N}=1$ $U(1)_R$ R-symmetry with the $\mathcal{N}=2$ $U(1)^{\mathcal{N}=2}_R \times SU(2)_R$ R-symmetry, that is seen as a global $U(1)$ symmetry from the $\mathcal{N}=1$ viewpoint. This symmetry is not seen in the UV Lagrangian we used and so must arise accidentally in the IR. We can do a consistency check by setting $\nu=1$ and noticing that (\ref{IndexE6}) indeed matches (\ref{IndexSU4}).

To summarize, the superconformal index of the $SU(4)$ gauge theory we introduced, after the removal of the free fields, matches that of the rank $1$ $E_6$ SCFT, at least to the order in $p q$ that we evaluated. The UV theory manifests a $U(1)_b \times SU(5)\times SU(2)$ global symmetry that is expected to enhance to the $E_6$ global symmetry of the SCFT. However, we do not see the $U(1)$ global symmetry that, together with $U(1)_R$, should form the $\mathcal{N}=2$ superconformal R-symmetry, implying that it must arise accidentally in the IR for our claim to be true. Given that the methods we are using rely on the correct identification of the superconformal R-symmetry, the appearance of accidental $U(1)$ symmetries is unfortunate. Nevertheless, the results here should still hold if this accidental $U(1)$ does not mix with the superconformal R-symmetry.   

\subsection*{Anomalies}

We can next consider the 't Hooft anomalies of the theory. These are also invariant under flows and so must match between the UV theory and the proposed IR theory. We have already noted that the conformal anomalies match, which was a factor in the conjecturing of this theory. The conformal anomalies can be evaluated using \eqref{Rtraces} and \eqref{aandc}, with $\gamma=0$. We find that:

\be
a = \frac{107}{48} = a_{E_6} + 25 a_{free} \quad , \quad c = \frac{77}{24} = c_{E_6} + 25 c_{free} ,
\ee
where $a_{E_6}, c_{E_6}$ and $a_{free}, c_{free}$ are the $a$ and $c$ central charges for the rank $1$ $E_6$ SCFT and a free chiral field respectively. The values of the former are given in \eqref{acE6} and those of the latter are given by:

\be
a_{free} = \frac{1}{48} \quad , \quad c_{free} = \frac{1}{24} .
\ee

We can next consider the anomalies involving the flavor symmetries. Here the only non-vanishing anomalies are of the form $Tr(U(1)^{\mathcal{N}=1}_R F^2)$ for $F$ a chosen flavor symmetry. This agrees with the $\mathcal{N}=2$ SCFT, as from there we expect the only non-vanishing anomalies to be of the form $Tr(U(1)^{\mathcal{N}=2}_R F^2)$, which indeed reduce to these ones. Here, from the $\mathcal{N}=2$ SCFT side, the only flavor symmetry is $E_6$ and we have that $Tr(U(1)^{\mathcal{N}=1}_R E^2_6) = -1$, see the appendix for details. Next we evaluate the anomalies of our theory involving the flavor symmetries that survive in the IR:

\be
Tr(U(1)^{\mathcal{N}=1}_R SU(2)^2) = 6 \times \frac{1}{2} \times (-\frac{1}{3}) = -1 ,
\ee 

\be
Tr(U(1)^{\mathcal{N}=1}_R U(1)^2_b) = 40 \times (-\frac{2}{3}) = -\frac{80}{3} ,
\ee 

\be
Tr(U(1)^{\mathcal{N}=1}_R SU(5)^2) = 8 \times \frac{1}{2} \times (-\frac{2}{3}) + 5 \times \frac{1}{3} = -1 ,
\ee  
where the second term in the last equation comes from the singlet field $M$. These should be equal to the result for the $E_6$ SCFT times the embedding index for the embedding. Consider an embedding of $H$ into $G$ and choose a representation $\boldsymbol{R_G}$ of $G$ that decomposes to $\sum \boldsymbol{R_H}_i$ representations of $H$ under the embedding. Then the embedding index is defined as:

\be
I_{H\rightarrow G} = \frac{\sum_i T_{\boldsymbol{R_H}_i}}{T_{\boldsymbol{R_G}}} ,
\ee
where $T_{\boldsymbol{R}}$ stands for the Dynkin index of the representation $\boldsymbol{R}$.

Using the embedding of $U(1)_b \times SU(5) \times SU(2)$ into $E_6$, we have that $I_{SU(2)\rightarrow E_6}=I_{SU(5)\rightarrow E_6}=1$, as well as:

\be
I_{U(1)_b\rightarrow E_6} = I_{U(1)_b\rightarrow SU(6)} I_{SU(6)\rightarrow E_6} = I_{U(1)_b\rightarrow SU(6)} = \frac{5 \times (\frac{2}{3})^2 + (-\frac{10}{3})^2}{\frac{1}{2}} = \frac{80}{3},
\ee
where we have used that $I_{SU(6)\rightarrow E_6}=1$. Overall, we find that all anomalies match.

The $\mathcal{N}=2$ $E_6$ SCFT also has anomalies involving the additional $U(1)$ commutant of $U(1)_R$ in the $\mathcal{N}=2$ R-symmetry, but as this symmetry is not visible in the UV Lagrangian, we cannot compare anomalies involving it.

\section{Duals for $R_{(2,2n+1)}$}
\label{sec:Rodd}

Having found a proposed dual for the rank $1$ $E_6$ SCFT, we can consider generalizing it to other cases. It turns out to be rather straightforward to generalize the previous Lagrangian to similar ones, that now appear to flow to the $R_{(2,2n+1)}$ family of theories, introduced in \cite{CD}, plus free chiral fields. We can also approach these using the anomalies, via the method we previously proposed, but as the generalization is rather straightforward, we shall approach it here from this angle instead. 

 Consider an $SU(2n+2)$ gauge theory with a chiral field in the antisymmetric representation, another in the conjugate antisymmetric representation, $2n+3$ chiral fields in the fundamental and $2n+3$ chiral fields in the antifundamental. This theory has non-anomalous global symmetry given by $SU(2n+3)^2 \times U(1)_a\times U(1)_b \times U(1)_c \times U(1)_R$, where the charges of the various fields are summarized in the table below. Particularly, it has a non-anomalous R-symmetry under which the antisymmetric chirals have free R-charge $\frac{2}{3}$, and the fundamental ones have R-charge $\frac{1}{3}$. For $n=1$, the matter content reduces to that of the previous theory.

\begin{center}
	\begin{tabular}{|c||c|c|c|c|c|c|c|}
		\hline
		Field & $SU(2n+2)$ & $SU(2n+3)_F$ & $SU(2n+3)_{\overline{F}}$ & $U(1)_a$ & $U(1)_b$ & $U(1)_c$ & $U(1)_{R}$\\
		\hline
		$A$ & $\boldsymbol{(n+1)(2n+1)}$ & $\boldsymbol{1}$ & $\boldsymbol{1}$ & $1$ & $0$ & $2n+3$ & $\frac{2}{3}$ \\
		
		$\overline{A}$ & $\boldsymbol{\overline{(n+1)(2n+1)}}$ & $\boldsymbol{1}$ & $\boldsymbol{1}$ & $-1$ & $0$ & $2n+3$ & $\frac{2}{3}$ \\
		
		$F$ & $\boldsymbol{2n+2}$ & $\boldsymbol{2n+3}$ & $\boldsymbol{1}$ & $0$ & $1$ & $-2n$ & $\frac{1}{3}$ \\
		
		$\overline{F}$ & $\boldsymbol{\overline{2n+2}}$ & $\boldsymbol{1}$ & $\boldsymbol{2n+3}$ & $0$ & $-1$ & $-2n$ & $\frac{1}{3}$ \\
		\hline
	\end{tabular}
\end{center}

We can repeat the dynamical analysis we did previously, but now for this more general set of theories. The expressions for generic $n$ are more complicated so we shall not write the explicit result though it remains qualitatively similar. Without any superpotential the superconformal R-symmetry is $U(1)^{sc}_R = U(1)_R + \gamma U(1)_c$, where $\gamma$ is a negative number for any $n$, as can be determined via a maximization. Like in the previous case, there is no mixing with $U(1)_b$ or $U(1)_a$ as the matter fields come in equivalent representations with opposite charges. With the R-symmetry found in this way all gauge invariant operators are above the unitarity bound and so it appears that this theory flows to an interacting SCFT. Specifically, the mesons $F \overline{F}$ have R-charge greater than the free R-charge $\frac{2}{3}$ due to the mixing.

We can next again consider the operator $Tr (A \overline{A} F \overline{F})$, given by the contraction of the $A$ and $\overline{A}$ fields that closes on the $SU(2n+2)$ adjoint multiplied by another $SU(2n+2)$ adjoint coming from the product of $F \overline{F}$, which is made invariant by taking the trace. This operator is in the $(\boldsymbol{2n+3}, \boldsymbol{2n+3})$ of the $SU(2n+3)_F \times SU(2n+3)_{\overline{F}}$ global symmetry, has charge $6$ under $U(1)_c$ and R-charge $2$ under $U(1)_R$. The last two imply that this operator is relevant with respect to the superconformal R-symmetry. As a result introducing it will initiate a flow leading to a new fixed point. The introduction of this operator breaks $U(1)_c$ and also breaks $SU(2n+3)_F \times SU(2n+3)_{\overline{F}}$ to the diagonal $SU(2n+3)$. As $U(1)_c$ is broken the R-symmetry is forced to be $U(1)_R$, under the assumption that there are no accidental $U(1)$ symmetries. Under this R-symmetry there are gauge invariant fields that have the free R-charge $\frac{2}{3}$, specifically, the mesons $F \overline{F}$. These should then decouple at the IR and we expect to get $(2n+3)^2$ free chiral fields plus a remaining fixed point. We claim that this remaining fixed point is the the $R_{(2,2n+1)}$ strongly coupled $\mathcal{N}=2$ SCFT. We next present evidence for this claim.

\subsection*{Index}

We shall start by studying the superconformal index of this theory. Naturally, it will be difficult to compute the index for arbitrary $n$. However, the lower order terms should be easy to compute for any $n$, which we shall do here. This gives some tests of the supersymmetry and global symmetry enhancements that are necessary for our proposal to hold. In the next subsection we shall consider the $n=2$ case in greater detail.

Like in the previous case, it is convenient to introduce additional singlet fields in order to remove the free fields at the IR. We shall again name these $M$ and $N$ and couple them to the mesons via the superpotential $W = M F \overline{F} + N F \overline{F}$. The charges of the two fields under the preserved global symmetry are shown in the table below. The F term relations of these fields then eliminate the mesons, $F \overline{F}$, from the chiral ring and thus remove them in the IR.   

\begin{center}
	\begin{tabular}{|c||c|c|c|c|c|}
		\hline
		Field & $SU(2n+2)$ & $SU(2n+3)$ & $U(1)_a$ & $U(1)_b$ & $U(1)_{R}$\\
		\hline
		$M$ & $\boldsymbol{1}$ & $\boldsymbol{(2n+3)^2-1}$ & $0$ & $0$ & $\frac{4}{3}$ \\
		
		$N$ & $\boldsymbol{1}$ & $\boldsymbol{1}$ & $0$ & $0$ & $\frac{4}{3}$ \\
		\hline
	\end{tabular}
\end{center}

We can next evaluate the index. For $n>2$, the terms up to order $p q$ should be uniform for any $n$. Generally the difference in $n$ here comes mostly from the baryons and the different number of degrees of freedom in the adjoint and antisymmetric representations. For $n>2$ these differences contribute at higher orders than $p q$ so we expect the index to be uniform. We find:

\bea
I & = & 1 + (p q)^{\frac{2}{3}} (2 + \boldsymbol{Ad}_{SU(2n+3)} +  \frac{b^2}{a} \boldsymbol{\Lambda^2}_{SU(2n+3)} + \frac{a}{b^2} \boldsymbol{\Lambda^{2n+1}}_{SU(2n+3)}) \nonumber \\ & - & p q (2 + \boldsymbol{Ad}_{SU(2n+3)} +  \frac{b^2}{a} \boldsymbol{\Lambda^2}_{SU(2n+3)} + \frac{a}{b^2} \boldsymbol{\Lambda^{2n+1}}_{SU(2n+3)}) + ... ,
\eea
where we use the fugacity $a$ for $U(1)_a$, as well as the notation $\boldsymbol{Ad}_{G}$ and $\boldsymbol{\Lambda^k}_{G}$ for the adjoint and rank $k$ antisymmetric representations of the group $G$, respectively. The index starts at order $(p q)^{\frac{2}{3}}$ as the mesons are killed by the flip fields. The first two terms are the contribution of the flip fields and the two last terms are gauge invariant operators build from two $F$ fields and one $A$ field and similarly for $\overline{F}$ and $\overline{A}$. The symmetric product of the mesons are again killed by the flip fields. The first two terms at order $p q$ come from the conserved currents that are preserved by the superpotential, where the broken conserved current were canceled against the operator that broke them. The last two terms come from $F^2 \Psi_{A}$ with a similar term from $\overline{F}$ and $\Psi_{\overline{A}}$, where $\Psi_{A}$ is the fermion in the chiral multiplet $A$. Again many terms involving the mesons are killed by the flip fields.

We note that the index form characters of $SO(4n+6)$ where $\boldsymbol{\Lambda^1}_{SO(4n+6)} \rightarrow \frac{b}{\sqrt{a}}\boldsymbol{\Lambda^1}_{SU(2n+3)} + \frac{\sqrt{a}}{b}\boldsymbol{\Lambda^{2n+2}}_{SU(2n+3)}$. In terms of $SO(4n+6)$ characters, the index reads:

\be
I = 1 + (p q)^{\frac{2}{3}} (1+\boldsymbol{Ad}_{SO(4n+6)}) - p q (1+\boldsymbol{Ad}_{SO(4n+6)}) + ... . \label{IndGen}
\ee

The first term can be identified with the moment map operators of the $U(1)\times SO(4n+6)$ symmetry of the $R_{(2,2n+1)}$ SCFT. The second term is then identified with the conserved currents of the $U(1)\times SO(4n+6)$ symmetry. Particularly, as we explicitly see them in the index, this is a strong indication that this symmetry enhancement occurs, again up to the caveat of possible mixing of the R-symmetry with accidental symmetries. Like in the previous case, we do not explicitly see in the UV Lagrangian the $U(1)$ commutant of the $\mathcal{N}=1$ $U(1)$ R-symmetry in the $\mathcal{N}=2$ R-symmetry. This implies that it should arise accidentally from the viewpoint of the UV theory. For our results to hold then, we must also assume that this $U(1)$ does not mix with the $\mathcal{N}=1$ $U(1)$ R-symmetry.

The index matches the known properties of the $R_{(2,2n+1)}$ SCFT. Specifically, the only relevant operators are the $U(1)\times SO(4n+6)$ moment map operators so the first order matches. The second order is a bit trickier. Here in the $R_{(2,2n+1)}$ SCFT we expect three contributions. One are the $U(1)\times SO(4n+6)$ conserved currents that we see in the index. The other two are the dimension three Coulomb branch operator and the conserved current for the additional $U(1)$ that is part of the $\mathcal{N}=2$ R-symmetry. The first contribute $\nu^3$ and the second $-1$ and so cancel if we set $\nu=1$. This is the same situation as we had before in the $n=1$ case, and stems from the fact that we do not see in the UV Lagrangian the commutant of $U(1)_R$ in the $\mathcal{N}=2$ R-symmetry.   

Finally, we want to consider the flavor $U(1)$. Specifically, we want to see how it is built in terms of $U(1)_a$ and $U(1)_b$. For that we need to identify the basic states charged under it in the $R_{(2,2n+1)}$ SCFT. These are known to be Higgs branch chiral ring generators in the spinor representation of $SO(4n+6)$, see the appendix for details. These contribute to the index as 

\be
\frac{(p q)^{\frac{(n+1)}{3}}}{\nu^{\frac{(n+1)}{2}}}(z \boldsymbol{S}_{SO(4n+6)} + \frac{1}{z}\boldsymbol{\overline{S}}_{SO(4n+6)}) \nonumber, 
\ee
where we are using the fugacity $z$ for the flavor $U(1)$, and the notation $\boldsymbol{S}_{SO(N)}$ for the spinor representation of $SO(N)$. Since here $N=4n+6$, this representation is always complex. To identify the $U(1)$ symmetry in the Lagrangian theory, we seek to find how these states originate in that frame. As the R-charge of these states seem to increase with $n$, it seems natural to identify them with similar states in the gauge theory, like baryons. Indeed, in the $SU(2n+2)$ gauge theory, we can build baryon type operators of the form $A^{n+1-m}F^{2m}$, and similar ones with $\overline{A}$ and $\overline{F}$. These all have R-charge $\frac{2(n+1)}{3}$, so they all contribute at the order expected to match the spinor Higgs branch chiral ring operators. Furthermore, under the decomposition of $SO(4n+6)$ to $U(1)\times SU(2n+3)$ we have that:

\be
\boldsymbol{S}_{SO(4n+6)} \rightarrow \sum^{n+1}_{m=0} (\frac{\sqrt{a}}{b})^{\frac{2n+3-4m}{2}} \boldsymbol{\Lambda^{2m}}_{SU(2n+3)},
\ee   
so to construct these operators from the gauge theory, we need states in antisymmetric representations of $SU(2n+3)$ of every rank. This is indeed provided by the $A^{n+1-m}F^{2m}$ and $\overline{A}^{n+1-m}\overline{F}^{2m}$ type operators. Looking at the flavor charges of all the $A^{n+1-m}F^{2m}$ type operators, we see that:

\be
\sum^{n+1}_{m=0} a^{n+1-m}b^{2m}\boldsymbol{\Lambda^{2m}}_{SU(2n+3)} = b^{\frac{2n+3}{2}}a^{\frac{2n+1}{4}} \sum^{n+1}_{m=0} (\frac{\sqrt{a}}{b})^{\frac{2n+3-4m}{2}} \boldsymbol{\Lambda^{2m}}_{SU(2n+3)} = b^{\frac{2n+3}{2}}a^{\frac{2n+1}{4}} \boldsymbol{S}_{SO(4n+6)},
\ee 
where we use fugacities to represent the $U(1)$ flavor charges, and have used the decomposition of $SO(4n+6)$ to $U(1)\times SU(2n+3)$ in the last equality. We can identify the right hand side with the $z \boldsymbol{S}_{SO(4n+6)}$ part. The $\overline{A}^{n+1-m}\overline{F}^{2m}$ type operators can then be identified with the $\frac{1}{z}\boldsymbol{\overline{S}}_{SO(4n+6)}$ part. From this we see that $z = b^{\frac{2n+3}{2}}a^{\frac{2n+1}{4}}$, which tells us how the $U(1)$ is spanned in terms of $U(1)_a$ and $U(1)_b$.

Unlike the $n=1$ case, for generic $n$, there are no known expressions for the full superconformal index of these theories that we can compare against. However, there are methods to compute various special limits of the index \cite{GRRY,GRRYpol,GR,GRR}. Many of these rely on specializations involving the additional R-symmetry present in an $\mathcal{N}=2$ SCFT, and so are inapplicable for our $\mathcal{N}=1$ Lagrangian theory where we do not explicitly see the Cartans of the $\mathcal{N}=2$ R-symmetry. There is one limit of the index, the so called Schur index \cite{GRRY,GRRYpol}, that can be taken even in our $\mathcal{N}=1$ Lagrangian theory \cite{RZCM}. This allows us to preform more precise tests. Naturally, it is challenging to do the computation, behind the orders already discussed, for any $n$. However, for low $n$, the calculation is quite manageable, specifically for $n=2$. As this case has some special features, we shall discuss it in detail later, but first we wish to turn to comparing 't Hooft anomalies.  

\subsection*{Anomalies}

Our next consistency check, is to match the 't Hooft anomalies of the $SU(2n+2)$ gauge theory to those of the $R_{(2,2n+1)}$ SCFT. The anomalies of the later are summarized in the appendix. Let us start with the conformal anomalies. For these, we find:

\bea
a & = & \frac{32n^2+58n+17}{48} = a_{R_{(2,2n+1)}} + (2n+3)^2 a_{free}, \nonumber \\  c & = & \frac{20n^2+40n+17}{24} = c_{R_{(2,2n+1)}} + (2n+3)^2 c_{free} ,
\eea
where we have used $a_{R_{(2,2n+1)}}$ and $c_{R_{(2,2n+1)}}$ for the $a$ and $c$ central charges of the $R_{(2,2n+1)}$ SCFT.

Next, we turn to the flavor anomalies. As the $R_{(2,2n+1)}$ theory is an $\mathcal{N}=2$ SCFT, all non-vanishing anomalies involve the $\mathcal{N}=2$ $U(1)$ R-symmetry, which in our case reduces to ones quadratic in the flavor symmetries and involving one insertion of the $\mathcal{N}=1$ $U(1)$ R-symmetry, as we do not explicitly see the other Cartan element in the $\mathcal{N}=1$ gauge theory. The values of these non-vanishing anomalies of the $R_{(2,2n+1)}$ theory are given in equation (\ref{AnomRodd}) in the appendix. It is straightforward to see that all anomalies that vanish in the $R_{(2,2n+1)}$ theory also vanish in the $SU(2n+2)$ gauge theory. For the non-vanishing anomalies in the gauge theory we find:

\be
Tr(U(1)^{\mathcal{N}=1}_R U(1)^2_a) = (2n+2)(2n+1) \times (-\frac{1}{3}) = -\frac{2(n+1)(2n+1)}{3} ,
\ee 

\be
Tr(U(1)^{\mathcal{N}=1}_R U(1)^2_b) = 2(2n+3)(2n+2) \times (-\frac{2}{3}) = -\frac{8(n+1)(2n+3)}{3} ,
\ee 

\be
Tr(U(1)^{\mathcal{N}=1}_R SU(2n+3)^2) = 2(2n+2) \times \frac{1}{2} \times (-\frac{2}{3}) + (2n+3) \times \frac{1}{3} = -\frac{(2n+1)}{3} ,
\ee  
where the second term in the last equation comes from the singlet field $M$. For the $SU(2n+2)$ symmetry we have that $I_{SU(2n+3)\rightarrow SO(4n+6)}=1$ so this anomaly matches. From the relation between the fugacities $z = b^{\frac{2n+3}{2}}a^{\frac{2n+1}{4}}$, $x=\frac{b}{\sqrt{a}}$, where we have defined $U(1)_x$ as the commutant of $SU(2n+3)$ in $SO(4n+6)$, we see that the symmetries are related as:

\be
U(1)_q = \frac{1}{n+1} U(1)_a + \frac{1}{2n+2} U(1)_b , \quad U(1)_x = \frac{2n+1}{4n+4} U(1)_b - \frac{2n+3}{2n+2} U(1)_a .
\ee

Therefore, we have that:

\be
Tr(U(1)^{\mathcal{N}=1}_R U(1)^2_q) = \frac{1}{(n+1)^2} \times (-\frac{2(n+1)(2n+1)}{3}) + \frac{1}{4(n+1)^2} \times (-\frac{8(n+1)(2n+3)}{3}) = -\frac{8}{3} ,
\ee

\bea
& & Tr(U(1)^{\mathcal{N}=1}_R U(1)^2_x) = \frac{(2n+3)^2}{4(n+1)^2} \times (-\frac{2(n+1)(2n+1)}{3}) + \frac{(2n+1)^2}{16(n+1)^2} \times (-\frac{8(n+1)(2n+3)}{3}) \nonumber \\ & = & -\frac{2(2n+3)(2n+1)}{3} .
\eea

The first one matches with the $U(1)$ anomaly (\ref{AnomRodd}) while the second one matches that expected from the $SO(4n+6)$ anomaly as:

\be
I_{U(1)_x\rightarrow SO(4n+6)} = \frac{(2n+3) \times (1)^2 + (2n+3) \times (-1)^2}{1} = 2(2n+3). 
\ee

\subsection{Other special cases}

As previously mentioned, there are some special cases for low $n$. One case is the $n=1$ case that we looked at initially. Here from the viewpoint of the $R_{(2,2n+1)}$ family, the operator in the spinor representation, that we previously found, contribute at the order of the moment maps. This leads to the enhancement of $U(1)\times SO(10)$ to $E_6$. The other physically interesting case is $n=2$. Here the spinor operator is a marginal operator. As the SCFT also possesses a dimension three Coulomb branch operator, there is an $\mathcal{N}=1$ only preserving conformal manifold \cite{RZCM}. In fact this theory is thought to have a dual description on the conformal manifold as an $\mathcal{N}=1$ conformal interacting gauge theory with the weak coupling point sitting on the conformal manifold\cite{RZCM}.

It is interesting to look at this case in a bit more detail. First we consider the full index, which here deviates from the general form of equation (\ref{IndGen}). The deviation is only through the additional spinor operator that we already previously determined so it is straightforward to write the index:

\bea
I & = & 1 + (p q)^{\frac{2}{3}} (2 + \boldsymbol{48}_{SU(7)} +  \frac{b^2}{a} \boldsymbol{21}_{SU(7)} + \frac{a}{b^2} \boldsymbol{\overline{21}}_{SU(7)}) + p q (a^3 + \frac{1}{a^3} + a^2 b^2 \boldsymbol{21}_{SU(7)} + \frac{1}{a^2 b^2}\boldsymbol{\overline{21}}_{SU(7)} \nonumber \\ & + & a b^4 \boldsymbol{\overline{35}}_{SU(7)} + \frac{1}{a b^4}\boldsymbol{35}_{SU(7)} + b^6\boldsymbol{\overline{7}}_{SU(7)} + \frac{1}{b^6} \boldsymbol{7}_{SU(7)} - \boldsymbol{48}_{SU(7)} -  \frac{b^2}{a} \boldsymbol{21}_{SU(7)} - \frac{a}{b^2} \boldsymbol{\overline{21}}_{SU(7)} - 2) + ... \nonumber \\  & = & 1 + (p q)^{\frac{2}{3}} (1+\boldsymbol{91}_{SO(14)}) + p q (z \boldsymbol{64}_{SO(14)} + \frac{1}{z}\boldsymbol{\overline{64}}_{SO(14)} -\boldsymbol{91}_{SO(14)}-1) + ... ,
\eea  
where like previously we have that $z=b^{\frac{7}{2}}a^{\frac{5}{4}}$, $\boldsymbol{14}_{SO(14)} \rightarrow \frac{b}{\sqrt{a}} \boldsymbol{7}_{SU(7)} + \frac{\sqrt{a}}{b} \boldsymbol{\overline{7}}_{SU(7)}$.

We can compare this against the index of the dual gauge theory proposed in \cite{RZCM}. As the gauge theory is related to the SCFT via going on the conformal manifold, some of the global symmetry of the SCFT is not seen in the gauge theory as it is broken by the marginal operators. First, the additional $U(1)$ Cartan of the $\mathcal{N}=2$ R-symmetry is broken on the conformal manifold. As the same symmetry is not exhibited by our Lagrangian, this is not an issue. In addition the $SO(14)$ global symmetry is broken to $U(1)\times SU(7)$ and only the non-abelian part and a combination of this $U(1)$ and the other $U(1)$ in the theory is preserved. When mapping to our theory, this corresponds to breaking $U(1)_a$, which can be implemented by setting $a = 1$ in the index. With this the indices match to order $p q$.   

Finally, we can match the Schur index of the $R_{(2,5)}$ theory with that of the $SU(6)$ gauge theory. In the gauge theory, the limit corresponding to the Schur limit of the $\mathcal{N}=2$ SCFT is taken by setting $p=q^2$. In this limit we find for the $SU(6)$ gauge theory index:
 
\be
I = 1 + 92 p^2 + 64(z+\frac{1}{z}) p^3 + 4266 p^4 + 5056(z+\frac{1}{z}) p^5 + 12(11086 + 143z^2 + \frac{143}{z^2}) p^6 + ...,
\ee
where for ease of computation, we have unrefined with respect to the $SU(7)$, and have also set $a=b^2$ so that $z=b^{\frac{7}{2}}a^{\frac{5}{4}} = b^6$. This index explicitly matches the Schur index of the $R_{(2,5)}$ theory evaluated using the results of \cite{GRRY} and the class S description of the $R_{(2,5)}$ theory in \cite{CD}.          

\section*{Acknowledgments}
We thank Shlomo Razamat for relevant discussions. GZ is supported in part by World Premier International Research Center Initiative (WPI), MEXT, Japan, by the ERC-STG grant 637844-HBQFTNCER and by the INFN.

\appendix

\section{Properties of the $R_{(2,2n+1)}$ family of SCFTs}
\label{AppA}

Here we shall summarize some properties of the family of strongly coupled $\mathcal{N}=2$ SCFTs known as the $R_{(2,2n+1)}$ theories. This family was introduced in \cite{CD}, and is comprised of strongly interacting $\mathcal{N}=2$ SCFTs that arise in dualities of $SU$ gauge groups with matter in the antisymmetric representation. They are described in class S by a compactification of the $A_{2n}$ $(2,0)$ theory on a sphere with three punctures, and have $U(1)\times SO(4n+6)$ global symmetry. 

The conformal anomalies of theories in this family are:

\be
a_{R_{(2,2n+1)}} = \frac{14n^2 + 23n + 4}{24} \quad , \quad c_{R_{(2,2n+1)}} = \frac{4n^2 + 7n + 2}{6} .
\ee    
 
We can also compute the anomalies involving the $U(1)\times SO(4n+6)$ flavor symmetry. The only non-vanishing ones are $Tr(U(1)^{\mathcal{N}=2}_R SO(4n+6)^2)$ and $Tr(U(1)^{\mathcal{N}=2}_R U(1)^2)$, where $U(1)^{\mathcal{N}=2}_R$ is the $\mathcal{N}=2$ superconformal $U(1)$ R-symmetry. Here for later convenience we shall instead use $Tr(U(1)^{\mathcal{N}=1}_R SO(4n+6)^2)$ and $Tr(U(1)^{\mathcal{N}=1}_R U(1)^2)$, where $U(1)^{\mathcal{N}=1}_R$ is the $\mathcal{N}=1$ superconformal $U(1)$ R-symmetry embedded in the $\mathcal{N}=2$ superconformal group. In terms of $U(1)^{\mathcal{N}=2}_R$ and $I_3$, the Cartan of the $SU(2)$ R-symmetry, it is given by\cite{BTW}:

\be
U(1)^{\mathcal{N}=1}_R = \frac{1}{3} U(1)^{\mathcal{N}=1}_R + \frac{4}{3} I_3. 
\ee

 The non-vanishing anomalies are then given by:

\be
Tr(U(1)^{\mathcal{N}=1}_R SO(4n+6)^2) = -\frac{(2n+1)}{3} \quad , \quad Tr(U(1)^{\mathcal{N}=1}_R U(1)^2) = -\frac{8}{3} . \label{AnomRodd}
\ee

Here the $U(1)$ is normalized such that the minimal charge is $1$. The anomaly involving the non-abelian part can be derived from the known flavor symmetry central charge of the $R_{(2,2n-1)}$ SCFT, $k_{SO(4n+6)}=4n+2$, using \cite{BTW}:

\be
Tr(U(1)^{\mathcal{N}=1}_R G^2) = -\frac{k_G}{6}.
\ee

 The 't Hooft anomaly involving the $U(1)$ is, to our knowledge, unknown, but it can be derived from dualities. Notably, the $R_{(2,2n+1)}$ participates in dualities relating to $SU(N)$ gauge theories with two antisymmetric hypermultiplets and four fundamental hypermultiplets \cite{CD}. Specifically, when $N=2n+1$ the dual is a gauging of $R_{(2,2n+1)}$ by a $USp(2n)$ gauge group with a fundamental hyper, while for $N=2n$ the dual is a gauging of $R_{(2,2n-1)}$ by a $USp(2n)$ gauge group with three fundamental hypers. It is straightforward to compute all the anomalies from the gauge theory side, and then by matching with the dual side, using the mapping of symmetries derived in \cite{BZ}, equation \eqref{AnomRodd} can be derived.

The $R_{(2,2n+1)}$ SCFT can also be described as a compactification of the $5d$ SCFT UV completion of a $5d$ $\mathcal{N}=1$ $USp(2n)$ gauge theory with $2n+3$ fundamental hypermultiplets \cite{BZ}. In this description the $SO(4n+6)$ is the symmetry rotating the $2n+3$ hypermultiplets while the $U(1)$ is the topological symmetry. The Higgs branch of this $5d$ SCFT was studied in \cite{FHMZ} and there it was found that, in addition to the $SO(4n+6)$ moment map operators, it has two Higgs branch chiral ring generators that are complex conjugates of one another, transform in the spinor representations of $SO(4n+6)$ and have charge $\pm 1$ under the $U(1)$. Their lowest component is a scalar in the $\boldsymbol{n+2}$ of $SU(2)_R$. From the $5d$ gauge theory description these come from 1-instanton configurations, which are expected to give a BPS operator whose lowest component is a scalar in the $\boldsymbol{n+2}$ of $SU(2)_R$, and, due to fundamental fermionic zero modes, also in the spinor representations of $SO(4n+6)$. As these are $1$-instanton configurations they carry charge $\pm 1$ under the topological $U(1)$. Note also that the $5d$ description naturally gives a normalization of the $U(1)$ where the minimal charge is $1$ and in that normalization the aforementioned states have this minimal charge.

\end{document}